\documentclass[onecolumn]{aa}
\usepackage{graphics, epsfig}
\newcommand{\1}{{\Omega_{\rm M} }}
\newcommand{\2}{{\Omega_{{\rm M0}} }}

\newcommand{\lab}{\label}
\begin{document}
\title{ Two viable quintessence models of the Universe: confrontation of theoretical predictions with observational data}
\author{M.Demianski \inst{1} \fnmsep \inst{2}\fnmsep \inst{3}, E. Piedipalumbo\inst{4}\fnmsep \inst{5}
C. Rubano\inst{4}\fnmsep \inst{5}, C.Tortora\inst{4}\fnmsep
\inst{5} }
\offprints{E.Piedipalumbo,\\
\email{ester@na.infn.it}} \institute{Institute for Theoretical
Physics, University of Warsaw,  Hoza 69, 00-681 Warsaw, Poland
\and Theoretical Astrophysics  Center, Juliane Maries Vej 30
DK-2100, Copenhagen, Denmark \and Department of Astronomy,
Williams College, Williamstown, Ma 01267, USA \and Dipartimento di
Scienze Fisiche, Universit\`{a} di Napoli Federico II, Compl.
Univ. Monte S. Angelo, 80126 Naples, Italy \and Istituto Nazionale
di Fisica Nucleare, Sez. Napoli, Via Cinthia, Compl. Univ. Monte
S. Angelo, 80126 Naples, Italy }
\date{Received / Accepted}
\titlerunning{Two viable quintessence models...}
\authorrunning{M.Demianski \& al.}
\abstract{ We use some of the recently released observational data
to test the viability of two classes of minimally coupled scalar
field models of quintessence with exponential potentials for which
exact solutions of the Einstein equations are known. These models
are very sturdy, depending on only one parameter - the Hubble
constant. To compare predictions of our models with observations
we concentrate on the following data: the power spectrum of the
CMBR anisotropy as measured by WMAP, the publicly available data
on type Ia supernovae, and the parameters of large scale structure
determined by the 2-degree Field Galaxy Redshift Survey (2dFGRS).
We use the WMAP data on the age of the universe and the Hubble
constant to fix the free parameters in  our models. We then show
that the predictions of our models are consistent with the
observed positions and relative heights of the first 3 peaks in
the CMB power spectrum, with the energy density of dark energy as
deduced from observations of distant type Ia supernovae, and {\ bf
with} parameters of the large scale structure as determined by
2dFGRS, in particular with the average density of dark matter. Our
models are also consistent with the results of the Sloan Digital
Sky Survey (SDSS). Moreover, we investigate the evolution of
matter density perturbations in our quintessential models, solve
exactly the evolution equation for the density perturbations, and
obtain an analytical expression for the growth index $f$. We
verify that the approximate relation $f\simeq \Omega_{{\rm
M}}^{\alpha}$ also holds in our models. } \maketitle
\section{Introduction}
Recent observations of the type Ia supernovae and CMB anisotropy
strongly indicate that the total matter-energy density of the
universe is now dominated by some kind of dark energy or the
cosmological constant $\Lambda$
(\cite{rie+al98,Riess00,per+al99,Riess04}). The origin and nature
of this  dark energy remains
unknown~(\cite{zel67,weinberg2,car}).\\  In the last several years
a new class of cosmological models has  been proposed. In these
models the standard cosmological constant  $\Lambda$-term is
replaced by a dynamical, time-dependent  component - quintessence
or dark energy - that is added to  baryons, cold dark matter
(CDM), photons and neutrinos. The equation of state of the dark
energy is given by $w_Q \equiv \rho_Q /p_Q$, $\rho_Q$ and $p_Q$
being, respectively, the pressure and energy density, and $-1 \leq
w_Q <0$, which implies a negative contribution to the total
pressure of the cosmic fluid.  When $w_Q =-1$, we recover a
constant $\Lambda$-term. One of the possible physical realizations
of quintessence is a cosmic scalar field, minimally coupled to the
usual matter action  (\cite{pe+ra88,cal+al98}). Such a field
induces dynamically a repulsive  gravitational force, causing  an
accelerated expansion  of the Universe, as recently discovered by
observations of distant  type Ia supernovae (SNIa)
(\cite{per+al99,rie+al98, Riess04}) and confirmed by WMAP
observations (\cite{spergel}). Accelerated expansion together with
the strong observational evidence that the Universe is spatially
flat ( \cite{deb+al00, spergel}) calls for an additional component
and quintessence could be responsible for the missing energy in a
flat Universe with a subcritical matter density. Quintessence
drives the cosmological expansion at late times and also
influences the growth of structure arising from gravitational
instability. Dark energy could cluster gravitationally only on
very large scales ($\geq 100$ Mpc), leaving an imprint on the
microwave background anisotropy (\cite{cal+al98}); on small
scales, fluctuations in the dark energy are damped and they do not
influence the evolution of perturbations in the pressureless
matter ( \cite{ma+al99}). On the scales we are considering in the
following, we assume that quintessence behaves as a smooth
component, so that in our analysis the formation of clusters is
due only to matter condensation, while the quintessence alters
only the background cosmic evolution. The leading candidates  for
the dark energy as suggested by fundamental physics include vacuum
energy, a rolling scalar field, and a network of slight
topological defects~(\cite{tu00,cl01}).  Moreover, an eternally
accelerating universe seems to be at odds with some formulations
of the string theory~(\cite{fischler01}). However, this is still
controversial. Actually in the last few years it has been
suggested that string theory could be compatible with the
presently considered cosmological models (see for example,
~\cite{hell01,Townsend03,gibbons01} and references therein). This
has stimulated a revival of interest in the exponential scalar
field quintessence. In different scenarios, such exponential
potentials can reproduce the present accelerated expansion of the
universe and some predict future deceleration. Moreover,  despite
the criticism that exponential potentials require fine tuning,
recently several authors  have pointed out that the degree of fine
tuning needed in this case is not greater than in other
scenarios~(\cite{cl01,ru+sc01,card1}). \\    In this work we show
that the two models of quintessence for which general exact
solutions of the Einstein equation are known are compatible with
recent  observational data, in particular with the power spectrum
of the temperature anisotropy of the CMBR, the observations of
type Ia  supernovae, and the parameters of the large scale
structure of matter distribution. As a first step we introduce a
new parametrization of quintessence models considered by Rubano et
al. and Rubano \& Scudellaro ~(\cite{mc,ru+sc01}), which avoids
the problem of branching of solutions highlighted in~Cardenas \&
al., (2002).
\\
Since the physical features of these models will be extensively
discussed in a forthcoming paper, in Sect. 2, we simply present
the basic equations of the quintessence models used in this paper.
In  Sect. 3 the linear perturbation equation is solved for the two
potentials and the growth of density perturbations is discussed;
in  Sect. 4 we show that predictions of our models are compatible
with the observationally established power spectrum of CMB
anisotropy, the SNIa data, and the results of estimates of the
average mass density of the universe from galaxy redshift surveys.
In  Sect. 5 we discuss the possibility of constraining the
equation of state of dark energy by age estimates of the universe.
Sect. 6 is devoted to final conclusions.
\section{Model description}
In this paper we consider two quintessence models, with a single and a
double exponential potential, for which exact analytic solutions are
available. The discussion of the physical properties as well as the
mathematical features of these models goes beyond the aims of this work;
they are presented in ~ Rubano \& al. (2004). Here we will only give the
basic relations.

\subsection{The single exponential potential}
We investigate spatially flat, homogeneous, and isotropic
cosmological models filled with two non-interacting components:
pressureless matter (dust) and a scalar field $\varphi$, minimally
coupled with  gravity. We first consider the potential introduced
in~ Rubano \& Scudellaro (2001),
\begin{equation}\lab{scal1}
V(\varphi) \propto \exp\left\{ -\sqrt{3\over 2}\varphi\right\}.
\end{equation}

For this potential the following substitution
\begin{eqnarray}\label{eq:aphi}
  a&=&{\left( u\,v \right) }^{\frac{1}{3}}\,,\\
  \varphi&=&-\left( {\sqrt{\frac{2}{3}}}\,\log (\frac{u}{v}) \right),
\end{eqnarray}
where $a$ is the scale factor, makes it possible to integrate the
Friedman equations exactly.  Setting as  usual $a(0)=0$ we have
\begin{eqnarray}
u&=& u_1 t,\\ v&=&v_1 t + v_2 t^3\nonumber,
\end{eqnarray}
where $u_1$, $v_1$ and $v_2$ are integration constants, so for
$a(t)$, we get
\begin{equation}
\lab{scalea} a^3(t)=u_1v_2t^2(t^2+{v_1\over v_2})\,.
\end{equation}
If by $H$ we denote the time dependent Hubble parameter then
\begin{equation}
\lab{scal2} H= {4\over{3t}}\left({{t^2+{v_1\over
{2v_2}}\over{t^2+{v_1\over v_2}}}}\right).
\end{equation}

To determine the integration constants $u_1$, $v_1$ and $v_2$ we
set the present time $t_0 = 1$. This fixes the time-scale
according to the (unknown) age of the universe. That is to say
that we are using the age of the universe, $t_0$, as a unit of
time. We then set $a_0 = a(1) = 1$, which is standard, and finally
$ H_0= H(1)$. Because of our choice of time unit it turns out that
our $H_0$ is not the same as the $H_0$  that appears in the
standard FRW model.
 The two conditions specified above allow one to express all the basic cosmological
parameters in terms of $H_0$. With these choices the whole history
of the universe has been {\it squeezed} into the range of time
$[0,1]$. Moreover this model is uniquely parametrized by $H_0$
only. Explicitly we have:
\begin{eqnarray}\label{eq:aHOMtime}
a^3(t)&=&{t^2\over 2}[(3H_0-2)t^2+4-3H_0],\label{eq:aHOMtime1}\\
H(t)&=&
{{2\left(2(3H_0-2)t^2+4-3H_0\right)}\over{3t\left((3H_0-2)t^2+4-3H_0\right)}},\label{eq:aHOMtime2}\\
\Omega_M&=&{{(4-3H_0)\left((3H_0-2)t^2+4-3H_0\right)}\over[2(3H_0-2)t^2+4-3H_0]^2},\label{eq:aHOMtime3}\\
\Omega_{\varphi}&=&{{(3H_0-2)t^2\left(4(3H_0-2)t^2+3(4-3H_0)\right)}\over
[2(3H_0-2)t^2+4-3H_0]^2}\label{eq:aHOMtime4}.
\end{eqnarray}
Therefore now the omega parameters of matter and the dark energy
are
\begin{equation}\label{eq:om_matternow}
\Omega_{M_0}\equiv\Omega_{M}(t=1)=\frac{2(4 -
3\,H_0)}{9\,H_0^2}\,,
\end{equation}

\begin{equation}\label{eq:omphinow}
\Omega_{\varphi_0}\equiv\Omega_{\varphi}(t=1)= \frac{\left(3\,H_0
-
  2\right)
\,\left(3 \,H_0 + 4 \right) }{9\,H_0^2}\,.
\end{equation}

The equation of state of dark energy evolves with time and the
parameter $w$ is given by
\begin{equation}\label{w}
w= -\frac{1}{2}  +
  \frac{ 3(3\,H_0 - 4) }{6(4 -
     3\,H_0) +
     8\left(3\,H_0 - 2
                \right)t^2}\,,
        \end{equation}
so that today we have
\begin{figure}
\centering \includegraphics[width=6 cm, height=4.5 cm]{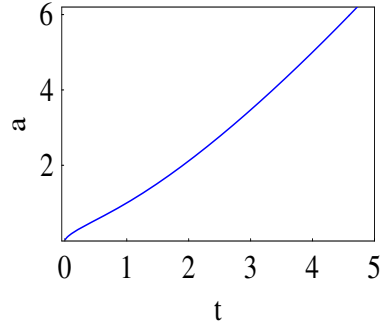}
        \caption{\small Time dependence of the scale factor $a$
for the single exponential potential, for
$H_0=0.97$.}\label{asinglet}
\end{figure}

\begin{figure}
\centering
        \includegraphics[width=6 cm, height=4.5 cm]{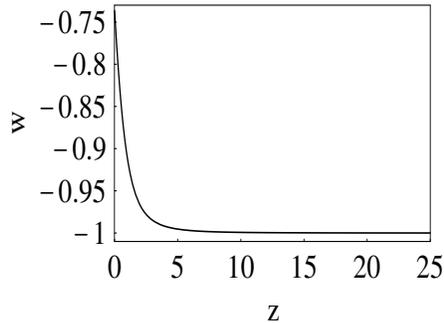}
        \caption{\small Redshift dependence of the parameter $w$
for the single exponential potential, for $H_0=0.97$.}
\label{fig:w}
\end{figure}

\begin{eqnarray}
  w_0 &=& -{8-3 H_0\over 4+3 H_0}\,.
\end{eqnarray}
In Fig. \ref{asinglet} we show the time evolution of the scale
factor while the redshift dependence of $w$ is plotted in Fig.
\ref{fig:w}. Asymptotically for $t\rightarrow \infty$, $a(t)\sim
t^{4/3}$ and therefore in this model the universe is eternally
accelerating and  possesses a particle horizon. The relation
between the dimensionless time $t$ and the redshift $z$ is given
by
\begin{equation}
\lab{scal4} z={2^{1\over
3}\over\left[t^2((3H_0-2)t^2+4-3H_0)\right]^{1\over 3}}-1.
\end{equation}
We shall compare the predictions  of the above model with a flat
cosmological model filled in with matter and the cosmological
constant. In this case, the redshift dependent Hubble parameter is
\begin{equation}
\lab{scal5} H(z)={\bar H}_0\sqrt{\2(1+z)^3+ (1-\2)}\,,
\end{equation}
where ${\bar H}_0$ is the standard Hubble constant. This class of
models has two free parameters, $\2$ and ${\bar H}_0$,  where
\[{\bar H}_0= {100\,h\,\,{\rm km} \, {\rm s}^{-1}\, {\rm Mpc}^{-1}}.\]
Let us assume that the age of the universe is
\[t_0=\gamma {\times}  1{\rm  Gy}=\,3.15\,10^{16}\,\gamma   \,  {\rm s}, \] where $\gamma$
is a constant to be determined by astronomical observations. With
this definition it is possible to  relate the value of $H_0$ to
the small $h={\bar H}_0/100$ of the standard FRW model. It turns
out that
\begin{equation}\label{eq:hubble-conversion}
  H_0= 0.1\,   h\,   \, \gamma.
\end{equation}
If we accept that $t_0=\,13.7\,$  Gy as given by the WMAP team
(\cite{spergel}) then $H_0=1.37 h$.
\subsection{The double exponential potential}

As the second example we consider the following double exponential
potential:
\begin{eqnarray}
\lab{scaldouble} V(\varphi)&=&\left( A\exp{({1\over
2}\sqrt{{3\over
2}\varphi})}-B \exp{(-{1\over 2}\sqrt{{3\over 2}}\varphi)}\right)^2\\
\nonumber &\equiv& A^2\exp{\left(\sqrt{{3\over
2}\varphi}\right)}+B^2 \exp{\left(-\sqrt{{3\over
2}}\varphi\right)}-2AB,
\end{eqnarray}
where $A$ and $B$ are constants. As we see, if we take $A,B>0$
this model contains in a certain sense an {\it intrinsic} negative
cosmological constant, but this does not lead to the re-collapsing
typical in such cases. It turns out that this is an eternally
expanding model, with alternate periods of accelerating and
decelerating expansion. In the following we denote
$\omega^2=3AB$.\\ For this model exact solutions of the Einstein
equations with scalar field exist, but their explicit mathematical
form is rather complicated. General properties of these solutions
are discussed in ~ Rubano \& al. (2004). Here we observe that if
we use the same parametrization as in the previous case, e.g.
taking the age of the Universe as a unit of time, the solution
will depend on only two parameters: $H_0$ and $\omega$. In order
to get physically acceptable values of $\Omega_{\varphi}$ (i.e
$\Omega_{\varphi}>0$) we have to take $\omega\leq 2$. However, in
our analysis we use a more restrictive range, $0\leq \omega \leq
1$, as we will discuss in the next sections. We will see however
that in this range of values of $\omega$ the matter density
parameter $\2$ is changing only slightly. In this case we obtain
the following expressions for $a$, $H$ and $\Omega_{\varphi}$,
which are the main quantities we need:
\begin{equation}\label{a2exp}
a^3(t)={{(3H_{0}\sin^{2}{\omega}-\omega\sin{2\omega})t^2- (3
H_{0}-2)\sin^2{\omega t}}\over
{2\sin^{2}{\omega}-\omega\sin{2\omega}}},
\end{equation}
\begin{equation}\label{H2aexp}
H={2\over
3}\left({t-\displaystyle{{\left(3H_0-2\right)\omega\sin{2 \omega
t}}\over {2(3 H_0\sin{\omega}-\omega \sin{2 \omega})}}\over
t^2-\displaystyle {{(3 H_0-2)\sin^2{\omega t}}\over {(3
H_0-2)\sin{\omega}-\omega\sin{2 \omega}}}}\right),
\end{equation}

\begin{eqnarray}\label{domegaphi}
&&\Omega_{\varphi}=\left[(\omega t\cos{\omega t}-\sin{\omega
t})^2+\omega^2\sin^{2}{\omega t}(t^2- {{(3H_0-2)\sin^2{\omega
t}}\over
{(3H_0\sin^2{\omega}-\omega\sin{2\omega})}})\right]{\times}\nonumber\\&&
{\left(3H_0-2\right)\over
(3H_0\sin^2{\omega}-\omega\sin{2\omega})\left[t- {{(3
H_0-2)\omega}\sin{2\omega
t}\over{2(3H_0\sin{\omega}-2\omega\cos{\omega})}}\right]^2}\,,
\end{eqnarray}
where $H_0$ and $\omega$ are constants. Since we consider a flat
model, $\Omega_{M}=1-\Omega_{\varphi}$. Note that when
$t\rightarrow \infty$, $\Omega_{\phi}\rightarrow {\rm const}$ and
in the generic case this constant is smaller than 1 and therefore,
when $t \rightarrow \infty$, $\Omega_{M}$ does not vanish, so at
the late stages of evolution of this model dark energy and matter
coexist. For large $t$, $a(t)\sim t^{2/3}$ though
$\Omega_{\varphi}\not=0$ and evolution of this model resembles the
matter-dominated phase of the standard FRW universe, hence in this
case the particle horizon does not appear. At the present epoch we
get
\begin{eqnarray}\label{eq:omega_phidnow}
&& \Omega_{\varphi_0}\equiv\Omega_{\varphi}(t=1)=\\
 &&  \frac{\left(  3\,H_0 -2\right) \,
    \left[3 H_0\left(\omega\sin{2\omega}-2\sin^{2}{\omega}\right)+2\,\omega\,(\sin{2\omega}-2\,\omega)\right]
      }{9\,{H_0}^2\,
    \left( \omega\,\sin{2\,\omega} -2
      \sin^2{\omega} \right)  }\,,\nonumber\\&&\nonumber\\
     &&\Omega_{M_0}\equiv\Omega_{M}(t=1)= {\left(2\over 3 H_0\right)}^2 \frac{\omega \,\sin{2\,\omega} +
  3\,H_0\cos^{2}{\omega} - 2  \,
     {\omega }^2\,}{
    \left( \omega \,\sin{2 \omega} -2\,\sin^2{\omega} \right) }\,.
\end{eqnarray}
When $\omega$ is small we obtain the following simple approximated
formula for $\Omega_{\varphi_0}$ and $\Omega_{M0}$
\begin{eqnarray}
  \Omega_{\varphi_0} &\cong& \frac{\left(3\,H_0 - 2\right)
      \,\left( 30\,
       \left(3\,H_0 + 4
                  \right)  -
      \left( 9\,H_0 + 4
         \right) \,
       {\omega}^2
      \right) }{27\,
    {H_0}^2\,
    \left( 10 -
      {\omega}^2
      \right) }\,,\label{omega_app}
\end{eqnarray}
\begin{eqnarray}
\Omega_{M_0}&\cong&\frac{2\,\left( 4\,
       \left(30 - \omega^2\right)  -
      3\,H_0\,\left(30 +  \omega^2 \right)
      \right) }{27\,H_0^2\,
    \left(10 -  \omega^2\right) }.\label{omegamat_app}
\end{eqnarray}

\begin{figure}
\centering{
        \includegraphics[width=5 cm, height=5 cm]{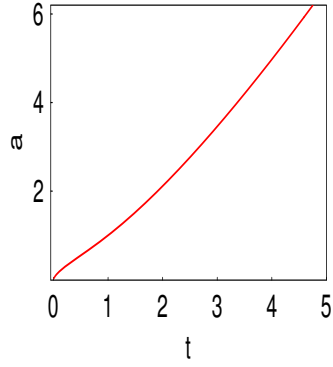}}
        \caption{\small Time dependence of the scale factor $a$
for the double exponential potential, for $H_0=0.97$, and
$\omega=0.1$.}\label{a_double}
 \end{figure}

For $\omega$ in the range $(0,1)$,  Eq. (\ref{omega_app}) shows
that the dependence of $\Omega_{\varphi_0}$ on $\omega$ can be
neglected. Actually, in analyzing the observational data we will
set $\omega=0.1$ . In the double exponential potential model the
relation between the dimensionless time $t$ and the redshift $z$
is given by

\begin{figure}
\centering{
        \includegraphics[width=6 cm, height=6 cm]{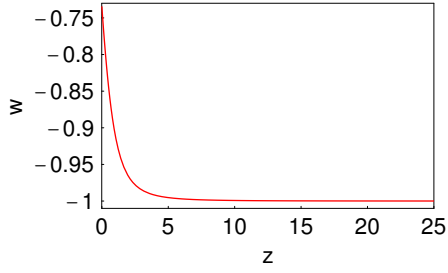}}
        \caption{\small Redshift dependence of the parameter $w$,
        for the double exponential potential, with $H_0=0.97$ and $\omega=0.1$.} \label{wosc1}
\end{figure}

\begin{eqnarray}\label{time-z}
  1+z &=&  \frac{
   {\left(\omega \,
           \sin{2\omega}-2\sin^{2}\omega
        \right)}^{1\over 3}}{\left[t^2\left(\omega \sin{2\omega}-3 H_0\sin^2{\omega}\right)+
        \left(3 H_0-2\right)\sin^2{\omega\,t}\right]^\frac{1}{3}}\,,
\end{eqnarray}
which for small values  of $\omega$ and $t$ can be simplified to
\begin{eqnarray}\label{t-zdoubleapp}
 1+z &\simeq& { {2^{1\over 3} t^{-{2\over 3}}}\over \left[4-3 H_0-{2\over
 15}\left(3\,H_0-2\right)\omega^2\right]^{1\over 3}}\,.
 \end{eqnarray}

 Fig. \ref{a_double} shows the time evolution of the scale factor
$a$.  As shown in Fig. \ref{wosc1} the equation of state for this
double exponential potential evolves with time.  The relation
between $H_0$ and the value of the Hubble constant in the FRW
model is the same as in the single exponential case.
\section{Growth of density perturbations}
The equation describing evolution of the CDM density contrast,
$\delta_M \equiv \delta \rho_M /\rho_M$, for perturbations inside
the  horizon, is (\cite{peeb80,ma+al99})
\begin{equation}
\lab{grow1} \ddot{\delta}_M+2H(t)\dot{\delta}_M-4\pi{\rm G }\rho_M
\delta_M=0,
\end{equation}
where the dot denotes the derivative with respect to time. In
Eq.~(\ref{grow1}) the dark energy enters through its influence on
the expansion rate $H(t)$. We shall consider Eq.~(\ref{grow1})
only in the matter dominated era, when the contribution of
radiation is really negligible.
\subsection{The single exponential case}
For the model with the single exponential potential described by
the Eq.~(\ref{scal1}), the differential equation~(\ref{grow1}) reduces to\\
\begin{eqnarray}\label{grow2}
 &&\frac{d^2\delta_M}{dt^2}+ \frac{4}{3t}\,\frac{\,\left[4 -3H_0 - 2(2 -
3\,H_0)\,t^2 \right]}{(4\,- 3\,H_0 - (2\ - 3\,H_0)\,t^2) }
\frac{d\delta_M}{dt}-\frac{2(4-3\,{H_0})\delta_M}{3t^2\left[ 4 -
3H_0\,- (2- 3\,H_0)t^2 \right]}=0.
\end{eqnarray}
Equation~(\ref{grow2}) is of Fuchsian type with 3 finite regular
singular points, and a regular point at $t\rightarrow \infty$;
i.e. it  is an hypergeometric equation, which has two linearly
independent  solutions, the growing mode $\delta_+$ and the
decreasing mode  $\delta_-$. Solutions of this equation can be
expressed in terms  of the hypergeometric function of the second
type $\ _2F_1$. We get
\begin{equation}
\lab{grow3} \delta_- \propto \frac{1}{t} \ _2F_1 \left[
-\frac{1}{2},\frac{1}{3},\frac{1}{6}; \left( {{3H_0-2}\over
{3H_0-4}}\right) \, t^2 \right],
\end{equation}
and
\begin{equation}
\lab{grow4} \delta_+ \propto t^{\frac{2}{3}}\,\ _2F_1 \left[
\frac{1}{3},\frac{7}{6},\frac{11}{6}, \left( {{3H_0-2}\over
{3H_0-4}} \right) \, t^2\right].
\end{equation}
In the linear perturbation theory the peculiar velocity field
${\bf v}$ is determined by the density contrast
(\cite{peeb80,padm93})

\begin{equation}
\lab{grow7} {\bf v} ({\bf x})= H_0 \frac{f}{4\pi} \int \delta_M
({\bf y}) \frac{{\bf x}-{\bf y}}{\left| {\bf x}-{\bf y} \right|^3}
d^3 {\bf y},
\end{equation}

where the growth index $f$ is defined as
\begin{equation}
\lab{grow8} f \equiv \frac{d \ln \delta_M}{d \ln a}\,,
\end{equation}
$a$ is the scale factor. According to our conventions, using only
the growing mode, we get

\begin{figure}[ht]
\centering{
\includegraphics[width=6 cm, height=6. cm]{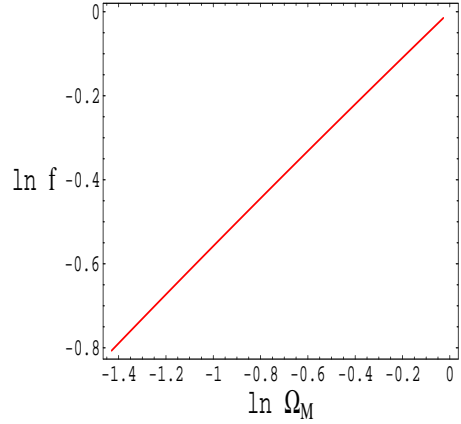}}
\caption{\small $\ln f$ versus $\ln \1$ in the single exponential
potential model.} \label{lnfomega}
\end{figure}
\begin{eqnarray}\label{grow9}
  f &=& \frac{4 - 3\,H_0 +
    \left( 3\,H_0 -2\right) \,t^2}
    {4 - 3\,H_0 +
    2\,\left(  3\,H_0-2 \right) \,
     t^2}\left\{\ _2F_1 \left[
\frac{1}{3},\frac{7}{6},\frac{11}{6}, \left( {{3H_0-2}\over
{3H_0-4}}\right) \, t^2\right]+\right.\\
\nonumber &&\left.\frac{7\,\left(
      3\,H_0 -2
      \right) \,t^2}{11\,
    \left(
      3\,H_0-4
      \right) }\ _2F_1 \left[
\frac{4}{3},\frac{13}{6},\frac{17}{6}, \left( {{3H_0-2}\over
{3H_0-4}}\right) \, t^2\right]\right\}\left({\ _2F_1 \left[
\frac{1}{3},\frac{7}{6},\frac{11}{6}, \left( {{3H_0-2}\over
{3H_0-4}}\right) \, t^2\right]}\right)^{-1}\,.
\end{eqnarray}

The growth index is usually approximated by $f \simeq \1^\alpha$.
For $\Lambda$CDM models, $\alpha \simeq 0.55$~(see
\cite{si+wa94,wa+st98,lokas}). In Fig.~{\ref{lnfomega}}, we show
the logarithm of $f$ as a function of the logarithm of $\1$ for
the single exponential potential. It turns out that $\alpha \simeq
0.57$ provides a good approximation to the model. In Fig.~6  we
see that $\alpha$ can be considered a constant during the late
stages of the universe evolution.
\begin{figure}
 \centering
\includegraphics[width=6 cm, height=5 cm]{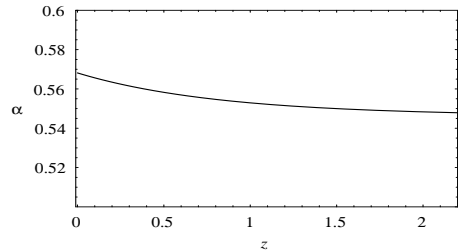}
 \caption{\small Redshift dependence of the exponent $\alpha$ in the
 single exponential potential model. We see that $\alpha$ can be
considered constant during the cosmic expansion.} \label{al}
\end{figure}

\subsection{The double exponential case }
For the model with double exponential potential described by
 Eq.~(\ref{scaldouble}), the differential equation~(\ref{grow1})
becomes more complicated, it assumes the form
\begin{eqnarray}
\label{groweq2} &&\frac{d^2\delta_M}{dt^2}+\frac{2}{3}\,\,\left[{2
t\left(3H_0 \sin^2\omega-\omega \sin{2 \omega } \right)+ \omega \,
 \sin (2 \omega t)
 \left( 2 - 3\,{H_0} \right)\over t^2\left(3H_0 \sin^2\omega-\omega
\sin{2\,\omega}\right) + \sin^2(\omega t) \left(2 - 3{H_0}
\right)}\right]\frac{d\delta_M}{dt}\\ \nonumber &&-{4\left[3
H_0\left(\omega^2-\sin^2{\omega}\right)+\omega\left(\sin{2
\omega}- 2\omega\right)\right]\over
 3 \left[2\left(3H_0 -2\right)\sin^{2}{\omega t}+2 t^2\left(\omega\sin{2 \omega}-
 3H_0 \sin^2{\omega}\right)\right]}\delta_M =0\,.
\end{eqnarray}
 Eq.~(\ref{groweq2}) does not admit exact analytic solutions.
However, since with our choice of normalization the whole history
of the Universe is confined to the range $t\in[0,1]$, and since we
choose $\omega \leq 1$, we can expand  the trigonometric functions
appearing in  Eq.~(\ref{groweq2}) in series around $t=1$,
obtaining an integrable differential equation, which is again a
hypergeometric equation.  For the growing mode we get
\begin{eqnarray}\label{growmode2}
&&\delta_+\propto t^{2\over 3}\ _2F_1 \left[
-\frac{1}{3},\frac{7}{6},\frac{11}{6}; {\left(2-3H_0\right)\,
t^2\over {2\left[3H_0-4- {2 \over
5}(H_0-2)\omega^2\right]}}\right]\,.
\end{eqnarray}

We use the growing mode $\delta_+$ to construct the growth index
$f$; according to  Eq.~(\ref{grow8})  we  obtain
\begin{equation}\label{fapp}
f= {A_1\over A_2},
\end{equation}
where
\begin{eqnarray}
&& A_1 = {105\over 4} (3H_0-4)^3 (2-3 H_0) t^2  \ _2F_1 \left[
-\frac{4}{3},\frac{16}{6},\frac{17}{6}; { \left({2-3H_0}\right)\,
t^2\over {2\left[3H_0-4-{2\over 5}(H_0-2)\omega^2\right]}}
\right]\\&& {\times}  \left(33\left[3H_0-4-{2\over
5}(H_0-2)\omega^2\right]\ _2F_1 \left[
-\frac{1}{3},\frac{7}{6},\frac{11}{6}; {\left(2-3H_0\right)\,
t^2\over {2\left[3H_0-4- {2 \over
5}(H_0-2)\omega^2\right]}}\right]\right)\,,\nonumber \\ && A_2 =
t^2 \omega^2 (3H_0-2)\left[\omega^2\left(176+39(H_0-4)H_0\right)+
(H_0-2)(3H_0-4)\right]\\ && +{5\over 2}(4-3 H_0)\left[(3H_0-2)t^2
+ 4 - 3H_0 \right]\,.\nonumber
\end{eqnarray}

\begin{figure}
\centering
        \includegraphics[width=7 cm, height=4.5 cm]{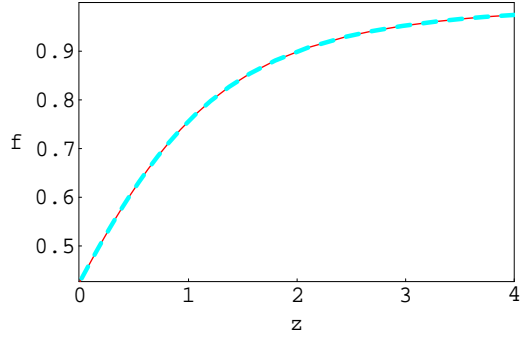}
        \caption{\small Behaviour of the exact and approximate growth index $f$
        for the double exponential potential model. The solid line is the exact
        function and the dashed one the approximate function.}\label{approx}
\end{figure}

 In
Fig.~\ref{approx} we see how accurate  our approximate $f$ is, in
comparison with that obtained by numerical integration of Eq.
(\ref{groweq2}).
 In this case the growth index is
approximated by $f\simeq\1^\alpha$, where $\alpha \simeq 0.57$.
\section{Observational data and predictions of our models}

\begin{figure}
\centering{
\includegraphics[width=7 cm, height=6.5 cm]{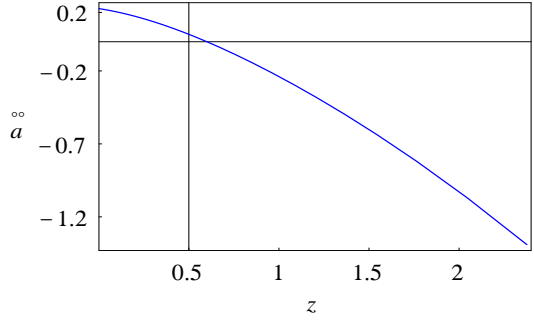}}
 \caption{\small Behaviour  of the second derivative of the scale
   factor
in the single exponential potential with
 $H_0=0.97$, $\Omega_{\varphi_0}=0.75$. Please note that the transition
from a decelerating to an accelerating expansion occurs close to
$z= 0.5 $,
 as predicted by recent observations of SNIa $z_t= 0.46\pm 0.13$ (\cite{Riess04}).}
\label{transition}
\end{figure}

The scalar field models of quintessence described in the previous
sections depend on at most two  arbitrary parameters that admit a
simple physical interpretation. The cosmological model with single
exponential potential of Eq. (\ref{scal1}) is uniquely
parametrized by the Hubble constant $H_0$ only, while the model
with double exponential potential of Eq. (\ref{scaldouble}) is
parametrized by $H_0$ and the frequency $\omega$. We have shown,
however, that the actual value of $\omega$  affects only slightly
the most important quantities as, for instance, the density
parameter $\Omega_{\varphi_0}$ (see Eq. (\ref{omega_app})).
Comparing predictions of this model with observations we will
assume that $\omega=0.1$. Using the WMAP data on the age of the
universe $t_{0}=13.7\pm 0.2$ Gy and on the Hubble constant ${\bar
H_{0}}=71\pm 5$ kms$^{-1}$Mpc$^{-1}$ (\cite{spergel}), from
Eq.(17) we get that our $H_{0}=0.97\pm 0.08$. Once $H_{0}$ and
$\omega$ are fixed our models are fully specified and  do not
contain any free parameters. In particular in the single
exponential potential model we have
$\Omega_{\varphi_0}=0.73\pm0.09$ and $w_{0}=-0.73\pm0.06$ and in
the double exponential potential model
$\Omega_{\varphi_0}=0.74\pm0.10$ and $w_{0}=-0.73\pm0.07$.  We can
now compare predictions of our models with available observational
data to test their viability. In Fig. \ref{transition} that with
these values of $H_0$ and $h$, the transition redshift from a
decelerating to an accelerating phase in the evolution of the
universe falls very close to $z=0.5$, in agreement with recent
results coming from the SNIa observations (\cite{Riess04}).
In the following we concentrate on three different kinds of
observations: namely the measurements of the anisotropy of the
CMBR by the WMAP team (\cite{Bennett03,spergel}), on the high z
supernovae of type Ia, and the observations of the large scale
structure by the 2dFGRS Team (\cite{haw02}).
\subsection{Constraints from CMBR anisotropy observations}
After the discovery of the cosmic microwave background radiation
by Penzias \& Wilson (1965) several experiments have been devoted
to measuring the  temperature fluctuations of CMBR. The recently
released  Wilkinson Microwave Anisotropy Probe (WMAP) data opened
a new epoch in  CMBR investigations, allowing strict tests of {\it
realistic} cosmological models. The WMAP data are powerful for
cosmological investigations since the mission was carefully
designed to  limit systematic errors, which are actually very low
(\cite{Bennett03}). The WMAP measured the power spectrum of the
CMBR temperature anisotropy, precisely determining positions and
heights of the first two peaks. It turns out that the separation
of the peaks depends on the amount of dark energy today, the
amount at the last scattering, and some averaged equation of
state. Assuming that the radiation propagates from the last
scattering surface up to now in such a way that the positions of
the peaks are not changed, the peaks appear at multipole moments
\begin{eqnarray}\label{picchi1}
  l_n &=& {\eta_0-\eta_{rec}\over c_s\eta_{rec}} n \pi \equiv n l_A,
\end{eqnarray}
where $\eta$ is the conformal time ( then $\eta_{rec}$ and
$\eta_0$ are respectively  its value at the recombination and
today), the speed of sound $c_s$ is assumed approximately constant
during recombination, and $l_A$ is the acoustic horizon scale. In
models with quintessence Eq. (\ref{picchi1}) should be modified
and rewritten as (\cite{doran, doran2, hu, hu01})
\begin{eqnarray}\label{picchi2}
  l_n &=&l_A (n-\zeta_n),
\end{eqnarray}
where $\zeta_n=\zeta+\Delta \zeta$, here $\Delta \zeta$ is a
general phase shift, whose form for the first three peaks is given
in (\cite{doran}), and
$\zeta=a_1(r_{rec})^{a_{2}}+\Omega_{\phi}^{rec}$, where
\begin{eqnarray}
&&r_{rec}={\rho_r(z_{rec})\over \rho_M(z_{rec}) },\\
&&\Omega_{\varphi}^{rec}=\eta_{rec}^{-1}\int^{\eta_{rec}}_{0}
\Omega_{\phi}(\eta) d\eta.
\end{eqnarray}

 The constants $a_1$ and $a_2$ are fit parameters also furnished
 in  Doran \& Lilley (2002), Doran \& al. (2002), and $\rho_r$ is the radiation density. It is possible to obtain
an analytic formula for $l_A$. Following Doran \&  Lilley and
introducing
\begin{equation}
 \langle w_0\rangle ={\int^{\eta_{rec}}_{0}
\Omega_{\varphi}(\eta) w(\eta)d\eta \over \int^{\eta_{rec}}_{0}
\Omega_{\varphi}(\eta) d\eta},
\end{equation}
we get
\begin{eqnarray}\label{lA}
  l_A &=& \pi c^{-1}_s\left[{F(\Omega_{\varphi_0},  \langle w_0\rangle)\over
  \sqrt{a_{rec}(1-\Omega_{\varphi}^{rec})}}
  \left\{1+ \left({\Omega_{r_0}\over a_{rec}(1-\Omega_{\varphi_0})}\right)^{1\over 2}
  +{\Omega_{r_0}\over 2 a_{rec}(1-\Omega_{\varphi_0})}\right\}-1\right]\,,
\end{eqnarray}

where
\begin{equation}
F(\Omega_{\varphi_0}, \langle w_0\rangle)={1\over 2}\int^1_0
dx\left(x+ {\Omega_{\varphi_0}\over  {1-\Omega_{\varphi_0}}}x^{1-3
 \langle w_0\rangle} + {\Omega_{r_0}(1-x) \over
{1-\Omega_{\varphi_0}}}\right)^{-{1\over 2}}\,,
\end{equation}
and $a_{rec}$ is the value of the scale factor at recombination.
Using  Eqs. (44), (\ref{lA}) and (\ref{picchi2}) in our single
exponential potential model, we find that $ \langle
w_0\rangle=-0.85$, and we get the following values of $l$
specifying the positions of the peaks \footnote{We note that $
\langle w_0\rangle=-0.85$ is fully compatible with the limit $w<
-0.78$ usually accepted for the present equation of state of dark
energy in quintessence models~(\cite{spergel}).}
\begin{eqnarray}\label{picchival}
  l_1 &=&225^{+ 11}_{- 12}\,, \\
  l_2 &=& 546^{+ 27}_{- 24}\,, \\
  l_3 &=& 830^{+ 42}_{- 29}\,.
\end{eqnarray}
 These values are consistent with the observed positions of the peaks
as measured by Boomerang (\cite{deb+al00}) and WMAP
(\cite{spergel}):
\begin{eqnarray}\label{picchiboom}
  l^{BOOM}_1 &=&222^{+ 14}_{- 14}\,, \\
  l^{BOOM}_2 &=& 539^{+ 21}_{- 21}\,,\\
l^{BOOM}_3 &=& 851^{+ 31}_{- 31}\,,
\end{eqnarray}

\begin{eqnarray}\label{picchiboom2}
  l^{WMAP}_1 &=&220.1^{+ 0.8}_{- 0.8}\,, \\
  l^{WMAP}_2 &=& 546^{+ 10}_{- 10}\,.
\end{eqnarray}

\begin{figure}
\centering{
        \includegraphics[width=8 cm, height=8 cm]{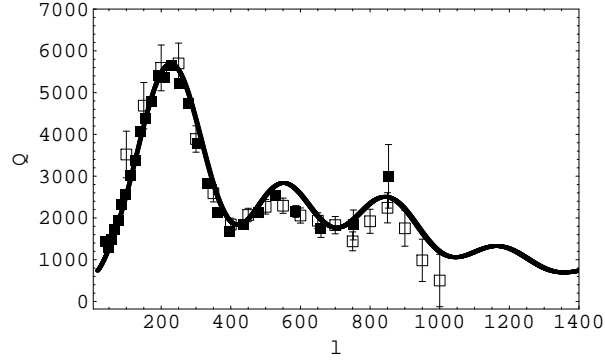}}
        \caption{\small Behaviour of the theoretical CMB anisotropy
        power spectrum for the single exponential
        scalar field model with $H_0=0.97$ compared with the data
 of Boomerang (empty boxes) and WMAP (filled boxes)}\label{picchis}
\end{figure}

In Fig. \ref{picchis} we plot the CMB anisotropy power spectrum
calculated for the single exponential potential model with
$H_0=0.97$, compared with the Boomerang data and WMAP
(\cite{deb+al00,spergel}).\\ It is also interesting to calculate
the relative height of the peaks. In particular  ${\cal H}_1$,
which is the first peak amplitude relative to the COBE
normalization ( that is, the height of the first peak is
normalized with respect to the COBE result at $l < 10$), ${\cal
H}_2$, the relative height of the second peak with respect to the
first, and ${\cal H}_3$, the amplitude of the third peak with
respect to the first. Using Camb (\cite{Antony}) with standard
input parameters to evaluate the CMB power spectrum, we obtained
${\cal H}_1=7.9$, ${\cal H}_2= 0.48$, and ${\cal H}_3=0.44$, while
a similar analysis performed on the Boomerang and Maxima data
gives ${\cal H}_1=7.6\pm 1.4$, ${\cal H}_2=0.45\pm 0.04$, and
${\cal H}_3=0.43\pm 0.07$.

\begin{figure}
\centering{
        \includegraphics[width=8 cm, height=8 cm]{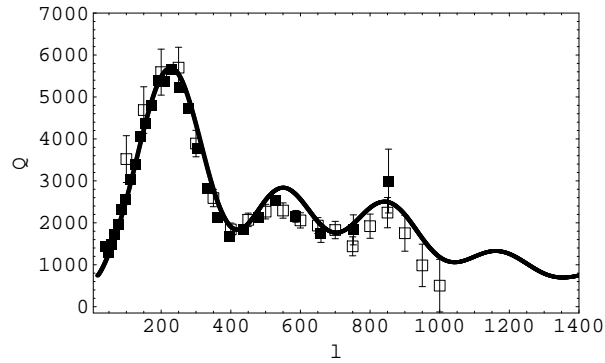}}
        \caption{\small Behaviour of the  CMB power spectrum for the double exponential
        potential model with $H_0=0.97$ and $\omega=0.1$ compared with the data
 of Boomerang (empty boxes) and WMAP (filled boxes).  }
        \label{picchid}
\end{figure}

As another complementary test we can use the WMAP result on the
CMB shift parameter $R=1.710\pm 0.137$, where (\cite{hu96})
\begin{eqnarray}
\label{R1}
  R &\equiv& H_0 \sqrt{\Omega_{M_0}}\int^{z_{ls}}_0{1\over H(z)}
  dz,
\end{eqnarray}

\begin{eqnarray}\label{R2}
 z_{ls}=1048 [1+0.00124 (\Omega_b h^2)^{-0.738}][1+g_1 (\Omega_{M_0}
h^2)^{g_2}]\,,
\end{eqnarray}
where $\Omega_b$ is the dimensionless baryon density and the
functions $g_1$ and $g_2$ are given in  Hu \& Sugiyama (1996).
According to  Eqs. (\ref{eq:aHOMtime2}), (\ref{eq:aHOMtime3}), and
(\ref{scal4}) $R$ can be written in terms of $t$ as
\begin{eqnarray}
  R &=& H_0 \sqrt{\Omega_{M_0}}\left(\left.{\cal F}(t)\right|_{t_0}-\left.{\cal
  F}(t)\right|_{t_{ls}}\right]),
\end{eqnarray}
where ${\cal F}(t)$ is the dimensionless comoving distance:
\begin{eqnarray}
  {\cal F}(t) &=& 3\,{\left(2t\right)}^{{1\over 3}}\,\frac{
    {\left( 1 -
        \frac{\left(
          3\,H_0 -2 \right) \,t^2}
          { 3\,H_0-4} \right) }
      ^{\frac{1}{3}}\,
    \ _2F_1\left[\frac{1}{6},
     \frac{1}{3},\frac{7}{6},
     \frac{\left(
         3\,H_0-2  \right)t^{2}}
        { 3\,H_0-4}\right]}{H_0\,
    {\left(t^2
        \left(
          3\,H_0-2  \right)+ 4-3 H_0
        \right) }^
     {\frac{1}{3}}}.
\end{eqnarray}
In our model, we obtain $\Omega_{M_0} h^2 =0.13\pm 0.03$,
$\Gamma=\Omega_{M_0} h=0.19^{+0.07}_{-0.07}$, so that
$R=1.71^{+0.09}_{-0.09}$, which is consistent with the WMAP value.
A similar analysis can be performed also for the model with the
double exponential potential (\ref{scaldouble}), even if the
calculations are more complicated and cannot be done by analytic
procedures only. The first complication is naturally due to the
presence of another parameter $\omega$. However, as shown in Eq.(
\ref{omega_app}), the final results depend only slightly on the
value of $\omega$ when picked from the acceptable range
$0<\omega\leq 1$. In analyzing the observational data we set
$\omega=0.1$.    Using Camb it is possible to calculate the
locations of the CMB power spectrum peaks also for this double
exponential potential model; we obtain
\begin{eqnarray}\label{picchivaldouble}
  l_1 &=&224^{+ 11}_{- 6}\,, \\
  l_2 &=& 543^{+ 24}_{- 17}\,, \\
  l_3 &=& 825^{+ 28}_{- 34}.
\end{eqnarray}
In Fig. \ref{picchid} we plot the CMB power spectrum resulting
from our model with the Boomerang and WMAP data. The evaluation of
the relative height of the peaks gives identical results as in the
single exponential potential case. In the same way we calculate
(this time numerically) the observable quantity R according to the
 Eqs. (\ref{R1}) and (\ref{R2}), and we obtain $\Omega_{M_0} h^2
=0.129\pm 0.05\pm 0.02$, $\Gamma=\Omega_{M_0}
h=0.19^{+0.07+0.02}_{-0.05-0.02}$, so that
$R=1.71^{+0.09}_{-0.09}$, which is consistent with the WMAP result
$R=1.70\pm 0.07$. We note that the second term in the errors takes
into account the effect of the {\it indeterminate}  value of
$\omega$.

\subsection{Constraints from recent SNIa observations}
In the recent years the confidence in type Ia supernovae as
standard candles has been steadily growing.  Actually it was just
the SNIa observations that gave the first strong indication of an
accelerating expansion of the universe, which can be explained by
assuming the existence of some kind of dark energy or nonzero
cosmological constant (\cite{Schmidt}).
 Since 1995 two teams of astronomers - the
High-Z Supernova Search Team and the Supernova Cosmology Project -
have been discovering type Ia supernovae at high redshifts. First
results of both teams were published by Schmidt \& al. (1998) and
Perlmutter \& al. (1999). Recently the High-Z SN Search Team
reported discovery of 8 new supernovae in the redshift interval
$0.3\leq z \leq 1.2$ and they compiled data on 230 previously
discovered type Ia supernovae (\cite{Tonry}).  Later Barris \& al.
(2004) announced the discovery of twenty-three high-redshift
supernovae spanning the  range of $z=0.34 - 1.03$, including 15
SNIa at $z\geq 0.7$~.\\  Recently Riess \& al. ~(2004) announced
the discovery of 16 type Ia supernovae with the Hubble Space
Telescope. This new  sample includes 6 of the 7 most distant ($z>
1.25$) type Ia  supernovae. They determined the luminosity
distance to these  supernovae and to 170 previously reported ones
using the same set  of algorithms, obtaining in this way a uniform
gold sample of type Ia supernovae containing 157 objects. The
purpose of this section is to test our scalar field quintessence
models by using  the best SNIa dataset presently available. As a
starting point we consider the gold sample compiled in Riess \&
al. (2001). To constrain our models we compare through a $\chi^2$
analysis the redshift dependence of the observational estimates of
the distance modulus, $\mu=m-M$, to their theoretical values. The
distance modulus is defined by
\begin{equation}
m-M=5\log{D_{L}(z)}+5\log({c\over H_{0}})+25, \label{eq:mMr}
\end{equation}
where $m$ is the appropriately corrected apparent magnitude
including reddening, K correction etc., $M$ is the corresponding
absolute magnitude, and $D_{L}$ is the luminosity distance in Mpc.
For a general flat and homogeneous cosmological model the
luminosity distance can be obtained through an integral of the
Hubble function $H$, as
\begin{eqnarray}\label{luminosity}
D_L (z) &=& (1+z)\int^{z}_{0}{1\over H(\zeta)}d\zeta\,.
\end{eqnarray}

\subsubsection{The single exponential potential}
\begin{figure}
\centering{
        \includegraphics[width=8cm, height=8.
        cm]{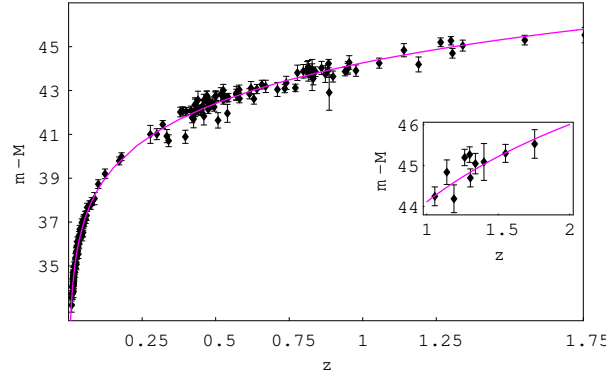}}
        \caption{\small Observational data of the gold sample of
SNIa (Riess et al. 2004)  fitted to our model with the single
exponential potential. The solid curve is the best fit curve with
$H_0=0.98^{+0.03}_{-0.05}$. It corresponds to $\Omega_{{\rm
M0}}=0.25_{-0.08}^{+0.05}$, and $h=0.65_{-0.04}^{+0.02} $. The
inner rectangle zooms in on the high redshift SNIa. }
        \label{riess_singlefit}
\end{figure}

For the single exponential potential model the luminosity distance
can be analytically calculated from  Eq. (\ref{luminosity}), using
the Hubble  function given in Eq. (\ref{eq:aHOMtime2}), and the
$z(t)$  relation as given by Eq. (\ref{scal4}).    The luminosity
distance can be represented in the following way:
\begin{eqnarray}
  D_L(t) &=& \left(1+z(t)\right)
  \left({3\sqrt{\pi}\,\, \Gamma[{7\over 6}]\over \Gamma[{2\over
  3}]}-
  3({2t\over {1+ t^2}})^{1 \over 3}\,\,
   _2F_1\left[{1\over 6},{1\over 3},{7\over 6},-t^2\right]\right).
\end{eqnarray}
Inverting the relation $z(t)$, we can construct $D_L(z)$, and
evaluate the distance modulus according to  Eq. (\ref{eq:mMr}).
Performing an $\chi^2$ analysis with the gold dataset of Riess \&
al. (2004) we obtain $\chi^2_{red}=1.15$ for 157 points, and as
best fit for $H_0$ the value $H_0=0.98^{+0.03}_{-0.05}$, which
corresponds to $\Omega_{{\rm M0}}=0.25_{-0.08}^{+0.05}$, and
$h=0.65_{-0.04}^{+0.02}$. These values are consistent with the
WMAP data analyzed above. In Fig. \ref{riess_singlefit} we compare
the best fit curve with the observational dataset.

\begin{figure}
\centering{
      \includegraphics[width=8 cm, height=8.
      cm]{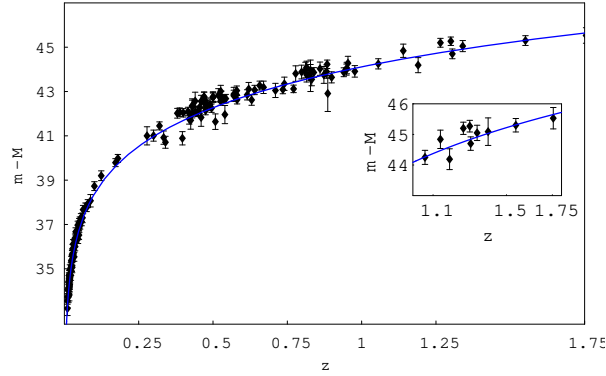}}
     \caption{\small Observational data of the gold sample of
SNIa (Riess et al. 2004 )  fitted to the double exponential
potential model. The solid curve is the best fit curve with
$H_0=0.99^{+0.03}_{-0.07}$. It corresponds to $\Omega_{{\rm
M0}}=0.25_{-0.06-0.02}^{+0.06+0.02}$, and
$h=0.65_{-0.04}^{+0.04}$. The last term in the error on
$\Omega_{{\rm M0}}$ quantifies the effect of the frequency
parameter $\omega$, when it changes from $\omega=0.1$ to
$\omega=0.9$. The inner rectangle zooms on the high redshift SNIa.
}
     \label{riess_doublefit}
\end{figure}

For the double exponential potential model
 we can perform the same analysis,
using  Eqs. (\ref{H2aexp}) and (\ref{time-z}). With our choice of
$\omega=0.1$ also in this case the $z(t)$ relation can be
inverted. Again, through  Eq. (\ref{luminosity}), which in this
case can be integrated only numerically, we construct the distance
modulus and perform the $\chi^2$ analysis on the gold data set. We
obtain $\chi_{red}^2=1.16$ for  157 data points,  and the best fit
value is $H_0=0.99^{+0.03}_{-0.07}$, which corresponds to
$\Omega_{{\rm M0}}=0.25_{-0.06-0.02}^{+0.06+0.02}$, and
$h=0.65_{-0.04}^{+0.04}$. The last set of  errors in $\Omega_{{\rm
M0}}$ quantifies the effect of the parameter $\omega$, when it
changes from $\omega=0.1$ to $\omega=0.9$. We noted that for this
potential the role of the high redshift supernovae is quite
important, and they change the value of $\chi^2_{red}$ from $1.1$,
if we consider the supernovae at $z\leq 1$, to
$\chi^2_{red}=1.17$, once we use the whole data set. This
circumstance confirms the necessity to increase the statistics of
data  at high redshifts to discriminate among different models. In
Fig. \ref{riess_doublefit} we compare the best fit curve with the
observational data.

\subsection{Constraint from galaxies redshift surveys}
  Once we know how the growth index $f$  changes with  redshift
and how it depends on $\Omega_{M}$ we can use the available
observational data to estimate the present value of $\Omega_{{\rm
M0}}$. The 2dFGRS team has recently collected positions and
redshifts of about 220,000 galaxies and presented a detailed
analysis of the two-point correlation function. They measured the
redshift distortion parameter $\beta=\displaystyle{f\over b}$,
where $b$ is the bias parameter describing the difference in the
distribution of galaxies and mass, and obtained that
$\beta_{|z\rightarrow 0.15}=0.49 \pm 0.09$ and $b=1.04 \pm 0.11$.
From the observationally determined $\beta$ and $b$ it is now
straightforward to get the value of the growth index at $z=0.15$
corresponding to the effective depth of the survey. Verde \&
al.~(2001) used the bispectrum of 2dFGRS galaxies, and Lahav \&
al. ~(2002) combined the 2dFGRS data with CMB data, and they
obtained
\begin{eqnarray}\label{bias}
 b_{verde}&=&1.04\pm 0.11\,,\\
 b_{lahav}&=&1.19\pm 0.09\,.
\end{eqnarray}

Using these two values for $b$ we calculated the value of the
growth index $f$ at $z=0.15$, we get respectively
\begin{eqnarray}\label{peculiar}
 f_1&=&0.51\pm 0.1\,,\\
 f_2&=&0.58 \pm 0.11\,.
\end{eqnarray}

\subsubsection{The single exponential potential}
Using Eq.(15) we express time $t$ through the redshift $z$ in
Eq.~(\ref{grow9}) and then substituting $z=0.15$ and the two
values of $f_1$ and $f_2$ we calculate $H_0$. Substituting the
thus obtained $H_0$ into Eq.(9) and setting $t=1$ we get
\begin{eqnarray}
 \2_1&=&0.26 \pm 0.1\,,\\
 \2_2&=&0.32\pm 0.13\,,\end{eqnarray}
which can be weighted to give the final value $\2=0.27\pm 0.09$.
Substituting the previously obtained relation between time $t$ and
the redshift $z$ and $H_0$ into Eq.(9) we calculate $\1(z=0.15)$
and we get
\begin{eqnarray}\label{omega_m}
 \1_1&=&0.32\pm 0.1\,,\\
 \1_2&=&0.39 \pm 0.13\,,
\end{eqnarray}
which gives the final value $\1(z=0.15)=0.35\pm 0.09$. It turns
out that this value is also fully compatible with the  independent
estimates derived, for instance, from the first data  release of
the SDSS (\cite{sloan1}).
\subsubsection{The double exponential potential}
Repeating the same procedure in the case of models with double
exponential potential (with $\omega=0.1$), we get
\begin{eqnarray}\label{omega_mnow2}
 \2_1&=&0.26 \pm 0.1\,,\\
 \2_2&=&0.31 \pm 0.11\,,
\end{eqnarray}
which can be weighted to give the final value $\2 =0.28\pm 0.07\pm
0.04$, where the last set of errors in $\2$ quantifies the effect
of $\omega$ when it changes from $\omega=0.1$ to $\omega=0.9$. As
in the previous case, using Eq.~(\ref{t-zdoubleapp}) we express
time $t$ through the redshift $z$ and using Eq.~(21) we calculate
$\1(z=0.15)$; we get
\begin{eqnarray}\label{omega_m2}
 \1_1&=&0.32 \pm 0.1\,,\\
 \1_2&=&0.38\pm  0.11\,,
\end{eqnarray}
which gives the final value $\1(z=0.15)=0.35\pm  0.07\pm 0.04$.

\subsection{Comparison with the standard $\Lambda$CDM model}
\begin{figure}
    \centering{
      \includegraphics[width=7 cm, height=6.5
      cm]{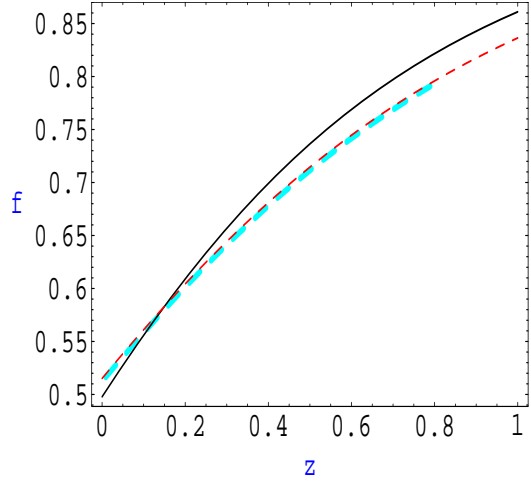}}
        \caption{\small The growth index $f$ in several cosmological
          models: the solid line corresponds to the $\Lambda$CDM
        model with $\2=0.25$. The dashed curves correspond to the two exponential
        potentials: the thin line is for the model with single
        exponential potential
        and the thick one is for the double
        exponential potential model.}
        \label{compare}\end{figure}

In the standard $\Lambda$CDM model we can write the growth index
as a function of the redshift $z$ and $\2$, obtaining:
\begin{eqnarray}\label{Lam}
  f_{\Lambda}&=&{\left( 1 + z \right)^{2} \,
    \ _2F_1\left[\frac{1}{3},1,\frac{11}{6},
     -\left( \frac{1 - \2}{\2} \right) \right]\over
 \ _2F_1\left[\frac{1}{3},1,\frac{11}{6},
  -\left( \frac{1 - \2}{\2\,{\left( 1 + z \right) }^3} \right) \right]}{\times}  \\ \nonumber
&& \left( \frac{\ _2F_1\left[\frac{1}{3},1,\frac{11}{6}, -\left(
\frac{1 -\2}{\2\,{\left( 1 + z \right) }^3} \right) \right]}{
{\left( 1 + z \right) }^2\,\
_2F_1\left[\frac{1}{3},1,\frac{11}{6}, -\left( \frac{1 -\2}{\2}
\right)\right]}  -  \frac{6\,\left( 1 -\2 \right) \, \
_2F_1\left[\frac{4}{3},2,\frac{17}{6},
 -\left( \frac{1 -\2}{\2\,{\left( 1 + z \right) }^3} \right) \right]}{11\,
 \2\,{\left( 1 + z \right) }^5\,
 \ _2F_1\left[\frac{1}{3},1,\frac{11}{6},
 -\left( \frac{1 -\2}{\2} \right) \right]}\right).
\end{eqnarray}
Applying the same procedure as in the case of the two quintessence
models we get
\begin{eqnarray}\label{omega_mlam}
 \2_1&=&0.22\pm  0.11\,,\\
 \2_2&=&0.28\pm  0.13\,,
\end{eqnarray}
and a final estimate of $\2=0.25 \pm 0.09$,  which is compatible
with the $\2$ obtained from independent measurements (see for
instance ~\cite{per+al99}) and our estimates. Therefore using only
the value of  $\2$ it is not possible to discriminate between the
models with the standard cosmological constant and the
quintessence models considered here. However, as is seen in
Fig.~\ref{compare}, independent measurements from large redshift
surveys at different depths can disentangle this degeneracy.

\section{The dark energy and the Hubble age }
It is known that the age estimates of the universe can in
principle independently constrain  the equation of state of the
dark energy, since the age of the universe, as well as the
lookback time - z relation, is a strongly varying function of $w$
(\cite{Krauss04, ferreras03}). In the approximation of constant
$w$, the age of a flat universe is defined by the formula:
\begin{eqnarray}\label{t-z1}
&& H_0t_0 = \int _{0}^{\infty }{dz \over{(1+z)} [\Omega_{m}(1+z)^3
+ \Omega_X(1+z)^{3(1+w)}]^{1\over 2}}\,,
\end{eqnarray}
where $\Omega_X$ is the fraction of the closure density in
material with the dark energy equation of state. The lookback time
is the difference between the present age of the universe and its
age at redshift z, and it is given by:
\begin{eqnarray}
\label{t-z2} && t(z) = t_H \int _{0}^{z }{dz \over{(1+z)}
[\Omega_{M0}(1+z)^3 +\Omega_X(1+z)^{3(1+w)}]^{1\over 2}},
\end{eqnarray}
where $t_H$ is the Hubble time. It is worth to note that since the
lookback time - z relation does not depend on the actual value of
$t_0$, it furnishes an independent and interesting cosmological
test trough age measurements, especially when it is applied to old
objects at high redshifts (\cite{lima}). In principle it is
possible that a model fits well the lookback time data and at the
same time gives a {\it wrong} value for $t_0$. For varying $w$, as
in the case of our models, the Eq. (\ref{t-z1}) can be rewritten
in a more general form as
\begin{eqnarray}
\label{t-z3} && H_0 t_0 = \int _{0}^{1}{da \over a F(a)},
\end{eqnarray}
where $ F(a)={H(a)\over H_0}$, while the lookback time relation
becomes
\begin{eqnarray}
\label{t-z4} && t(z) =t_H \int _{0}^{a }{dy \over y F(y)}.
\end{eqnarray}
The purpose of this section is to evaluate the Eqs. (\ref{t-z3})
and (\ref{t-z4}) for our quintessence models in order  to
constrain the value of the parameter $w$. Before going further we
note that this analysis is at the same time simple and
particularly interesting in the context of our parametrization,
which is based on the choice of the present age of the universe as
a unit of time.
\subsection{The single exponential model}
As a starting point we note that since the age of the universe has
been set equal to unity ($t_0=1$), the dimensionless quantity $H_0
t_0$ which appears in Eq. (\ref{t-z3}) is simply $H_0$. This means
that  $H_0$ is the only parameter that enters in the observational
quantities. From our previous analysis we obtained $H_0=0.97$,
which agrees with the limit of $H_0t_0 <1.1$ obtained in Krauss
(2004). Let us start with the lookback time - z relation, which
has the form
\begin{equation}\label{t-zexp1}
t(z,H_0)= {1\over \sqrt{2}} \left\{1 + \frac{1}{2 - 3\,H_0}
\left[2- \frac{\sqrt{9\,H_0^2\,{\left( 1 + z \right) }^3
+8\,z\left(z^2+3z+3\right)\left(2-3\,H_0\right) }}{ \left( 1 + z
\right)^{\frac{3}{2}}}\right]\right\}^{1\over 2}.
\end{equation}
Using the relation
\begin{eqnarray}
\Omega_M&=&{{(4-3H_0)\left((3H_0-2)t^2+4-3H_0\right)}\over[2(3H_0-2)t^2+4-3H_0]^2}\,,
\end{eqnarray}
we obtain $t=t(z,\Omega_{M})$ as
\begin{eqnarray}\label{time}
t(z,\Omega_M) &=&{\sqrt{2}\over (1+z)\sqrt{\sqrt{1+8 \Omega_M
}-\left(1+2 \Omega_M\right)}}\\ \nonumber && \left\{\sqrt{1+8
\Omega_M}\left(1+z\right)^{2}-\left(1+4
\Omega_M\right)\left(1+2z\right)+z^2\left(1-4
\Omega_M\right)\right.\\
\nonumber &&\left.-\sqrt{2\left(1+z\right)}\left[\sqrt{1+8
\Omega_M}\left(1+\left(1+4
\Omega_M\right)\left(3z+3z^2+z^3\right)\right)\right.\right.\\
\nonumber &&\left.\left.-\left(1+4 \Omega_M
\right)-z\left(3+3z+z^2\right)\left(1+8
\Omega_M\left(1+\Omega_M\right)\right)\right]^{1\over
2}\right\}^{1\over 2}\,.
\end{eqnarray}

\begin{figure}
\centering{
\includegraphics[width=7 cm, height=4.5 cm]{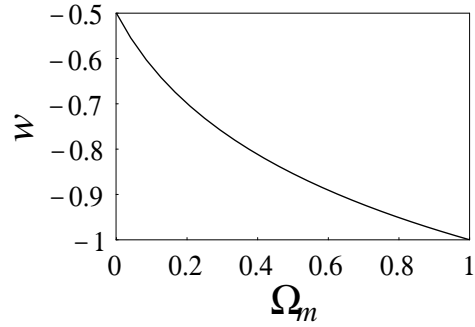}}
\caption{\small The curve $t(w,\Omega_M)=1$.}\label{t-wOm}
\end{figure}
If we impose, according to our assumptions, that $t(0,\Omega_M)=1$
we obtain  $\Omega_{M0}=\displaystyle
1-{\left(3H_0-2\right)\left(3H_0+4\right)\over 9H_0^2}$, which,
for $H_0=0.97 \pm 0.08$ gives $\Omega_{M0}=0.26\pm 0.05$. In a
similar way we can construct the {\it surface} $t(w,\Omega_M)$,
eliminating z between  Eq.(\ref{t-zexp1}) and (13). The curve
$t(w,\Omega_M)=1$ determines the physically acceptable values of
parameters in the plane $w -\Omega_M$, as shown in
Fig.\ref{t-wOm}.
\begin{figure}
\centering
\includegraphics[width=6 cm, height=6 cm]{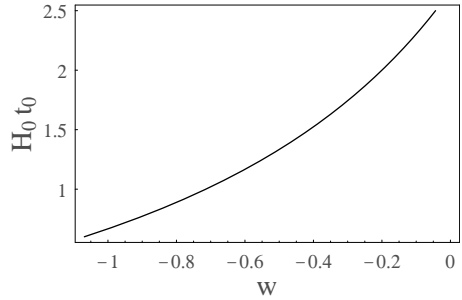}
\caption{\small Dependence of the dimensionless quantity $H_0 t_0$
on  $w$ for the single exponential model.}\label{hoto}
\end{figure}

In Fig.\ref{hoto}  we show the dependence of the dimensionless
quantity $H_0 t_0$ on the value of the  parameter $w$.
\subsection{The double exponential potential}
\begin{figure}
\centering{
\includegraphics[width=6 cm, height=5.5 cm]{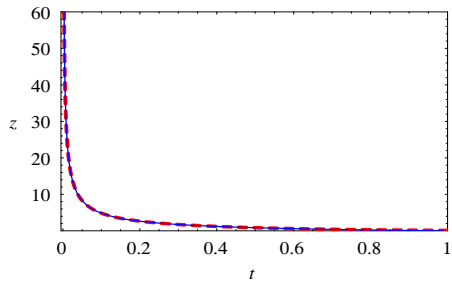}}
\caption{\small Behaviour of the exact and approximate $t-z$
relation for the double exponential model. The solid line is the
exact function and the dashed one the approximate
function.}\label{time-zdouble}
\end{figure}

For the model with double exponential potential the procedure
outlined for the single exponential potential becomes more
complicated from the computational point of view. Actually the
$z-t$ relation in  Eq.(\ref{time-z}) is rather involved and cannot
be exactly inverted. However, since $t\in (0,1)$ and for small
values of $\omega$, it can be simplified to the invertible form of
Eq. (\ref{t-zdoubleapp}), by means of a series expansion. In
Fig.\ref{time-zdouble} we compare the  exact result with the
approximate one.

Once $t(z)$ is known we can apply the same procedure as in the
single exponential case: it is possible to construct the {\it
surface} $t(w,\Omega_M)$, eliminating z between the  $w(t)$, and
$\Omega_{M}$. The curve $t(w,\Omega_M)=1$  determines the
physically possible values of parameters in the plane $w
-\Omega_M$.
\begin{figure}
\centering{
\includegraphics[width=6 cm, height=4.5 cm]{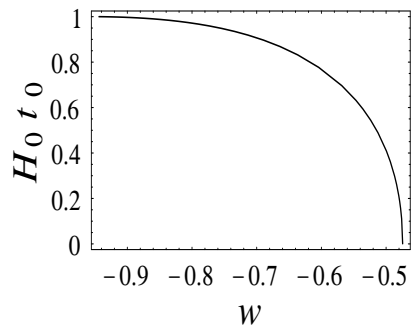}}
\caption{\small Dependence of the dimensionless quantity $H_0 t_0$
on $w$, for the model with double exponential
potential.}\label{hotodouble}
\end{figure}

In Fig.\ref{hotodouble}  we show the dependence of the
dimensionless quantity $H_0 t_0$ on the value of  the parameter
$w$.
\section{Conclusions}
We tested the viability of  two classes of cosmological models
with scalar field models of quintessence with  exponential
potentials. To compare predictions of our models with
observational  data we used  three different types of
observations: namely, the power spectrum of the CMB temperature
anisotropy and in particular positions and heights of the peaks,
the high-redshift SNIa data  compiled in Riess \& al. (2004) (gold
data set), and the large survey of galaxies by the 2dFGRS team
(\cite{haw02}) and in particular their estimate of the average
density of dark matter.  We showed that predictions  of our models

\begin{table*}[ht]
\begin{center}
\caption{The basic cosmological parameters derived from our models
are compared with those presented by the  WMAP team. The position
of the third peak $l_3$ has been determined by the Boomerang team.
The last set of  errors in the case of the double exponential
potential quantifies the effect of the parameter $\omega$, when it
changes from $\omega=0.1$ to $\omega=0.9$. }

\label{table-1_CMB} \vspace{0.5cm}
\begin{tabular}{|c|c|c|c|}
  \hline
 & $  \exp\left\{ -\sqrt{3\over 2}\varphi\right\}$ &  $ \left( A\exp{({1\over 2}\sqrt{{3\over2}\varphi})}-B \exp{(-{1\over 2}\sqrt{{3\over 2}}\varphi)}\right)^2$ & WMAP\\
   \hline
  $ \Omega_{M_0}h^2$& $0.13\pm 0.03$& $0.13\pm 0.03\pm 0.04$&$0.135\pm 0.008$\\
   \hline
    $ \Omega_{\varphi_0}$& $0.75\pm 0.08$& $0.76\pm 0.07\pm0.04$& $0.73\pm 0.04$\\
    \hline
 $ w_{0}$& $-0.76\pm 0.05$ &$ -0.74\pm 0.05\pm0.03$ &$w <
    -0.78$
    \\
    \hline
    $ \Gamma$& $0.19^{+0.07}_{-0.07}$ &$ 0.19^{+0.07+0.02}_{-0.05-0.02}$ &$ 0.19\pm 0.02$
    \\
 \hline
    $R$& $1.71^{+0.09}_{-0.09}$ &$1.71^{+0.09}_{-0.09}$ &$1.71\pm 0.137$
    \\
\hline
    $l_1$& $225^{+ 11}_{- 12}$ &$ 224^{+ 11}_{- 6}$ &$220.1^{+ 0.8}_{- 0.8}$
    \\
    \hline
    $l_2$& $546^{+ 27}_{- 24}$ &$ 543^{+ 24}_{- 17}$ &$546^{+ 10}_{- 10}$
    \\
\hline
    $l_3$& $830^{+ 42}_{- 29}$ &$ 825^{+ 28}_{- 34}$ &$851^{+ 31}_{- 31}$
    \\
\hline
\end{tabular}
 \end{center}
\end{table*}

are fully compatible with the  recent observational data. However
the models considered here, despite  being fully compatible with
the present day observational data have very different behavior on
long time scales. The model with the single exponential potential
will eternally accelerate, and though for sufficiently large t its
evolution is dominated by dark energy with
$\Omega_{\varphi}\rightarrow 1$ and $\Omega_{M}\rightarrow 0$ the
scale factor increases only as a power law $a(t)\sim t^{4/3}$ and
not exponentially. In this model there is a particle horizon. The
model with the double exponential potential has a different
asymptotic $(t\rightarrow \infty)$ behavior. First of all, though
$a(t)\rightarrow \infty$ as $t \rightarrow \infty$, $\Omega_{M}$
does not decrease to zero, so for large time matter and dark
energy coexist. For sufficiently large t the scale factor grows
like in a FRW  matter-dominated flat model $a(t)\sim t^{2/3}$. In
this model, the particle horizon does not form and asymptotically
( for large $t$) the expansion of the universe is decelerating. It
also turns out that the available observational data cannot
discriminate between the quintessence models considered here and
the $\Lambda$CDM model. New data coming from high redshifts
observations could remove this degeneracy.
\section*{Acknowledgments}
This work has been financially supported in part by the M.U.R.S.T.
grant PRIN ``DRACO.", the Polish Ministry of Science grant
1-P03D-014-26, and by EC network HPRN-CT-2000-00124.

\end{document}